\documentclass[aps,twocolumn,pra,superscriptaddress,showpacs,tightenlines]{revtex4-1}
\usepackage{amsmath}
\usepackage{graphicx}
\usepackage{color}
\usepackage{amsfonts}
\usepackage{CJK}

\begin{document}
\title{Photon blockade in quadratically coupled optomechanical systems}
\author{Jie-Qiao Liao}
\affiliation{CEMS, RIKEN, Saitama 351-0198, Japan}
\author{Franco Nori}
\affiliation{CEMS, RIKEN, Saitama 351-0198, Japan}
\affiliation{Department of Physics, The University of Michigan, Ann Arbor, Michigan 48109-1040, USA}

\date{\today}
\begin{abstract}
We study the steady-state photon statistics of a quadratically coupled optomechanical cavity, which is weakly
driven by a monochromatic laser field. We examine the photon blockade by evaluating the second-order
correlation function of the cavity photons. By restricting the system within the zero-, one-, and two-photon
subspace, we obtain an approximate analytical expression for the correlation function. We also numerically
investigate the correlation function by solving the quantum master equation including both optical and mechanical dissipations.
The results show that, in the deep-resolved-sideband and single-photon strong-coupling regimes,
the single-photon resonant driving will induce a photon blockade, which is limited by the
thermal noise of the mechanical environment.
\end{abstract}
%42.50.Pq Cavity quantum electrodynamics; micromasers
%42.50.Ar Photon statistics and coherence theory
%42.50.Wk Mechanical effects of light on material media, microstructures and particles
%07.10.Cm Micromechanical devices and systems
\maketitle

\section{Introduction}

The realization of strong photon correlations at the few-photon level has become an interesting and
important research topic in quantum optics~\cite{Ciuti2013,Imamoglu1997,Kimble2005,Faraon2008,Lang2011,Reinhard2012,Peyronel2012,Huang2013,Liao2010,Ferretti2010,Miranowicz2013}.
The significance of this subject is mainly motivated by the
considerable applications of correlated photons to the foundations of quantum theory as well as in
quantum information science. So far, much effort has been devoted to the creation of correlated photons in
various physical systems such as cavity-QED~\cite{Kimble2005,Faraon2008,Lang2011,Reinhard2012,Peyronel2012}
and Kerr-type nonlinear cavities~\cite{Liao2010,Ferretti2010,Miranowicz2013}. In particular, recent
attention has been paid to the generation of photon correlations in optomechanical
systems~\cite{Rabl2011,Liao2013,Liu2013,Marquardt2013,Ludwig2012,Stannigel2012,Lukin2013,Xu2013,Lu2012,Savona2013}.
It has been shown that \emph{linear} optomechanical couplings can cause a photon blockade in the combined single-photon strong-coupling
and resolved-sideband regimes, and that this photon blockade is modulated by the phonon sidebands. However, the photon blockade in
\emph{quadratically} coupled optomechanical systems has not been studied.

In a conventional photon blockade, it is believed that the optical nonlinearity in the eigenenergy spectrum is
the key element for obtaining the photon correlation. For example, in the Jaynes-Cummings (JC) system and
the Kerr-type nonlinear cavity, the optical nonlinearity takes the form $\pm\alpha\sqrt{1+ \beta s}$ and $\chi s^{2}$,
respectively, where $\alpha$ and $\beta$ are the parameters of the JC system, $\chi$ is the Kerr parameter, and $s$ is the photon number.
In quadratically coupled optomechanical systems, there is an optical nonlinearity of type $\omega_{M}\sqrt{1+4g_{0}s/\omega_{M}}$ [$g_{0}$ and $\omega_{M}$, cf. Eq.~(\ref{rotatingrepH})]
in the eigenenergy spectrum~\cite{Agarwal2008,Bhattacharya2013,Liao2013B}. When the single-photon optomechanical coupling
is strong enough, the optical nonlinearity could be used to create photon correlations.
Inspired by this feature, in this paper we study the photon blockade in a \emph{quadratically} coupled optomechanical
cavity. Concretely, we analytically and numerically study the steady-state photon statistics of the quadratic optomechanical cavity. By
examining the second-order correlation function, we clarify the photon blockade in this system by answering
the following three questions:

(1) What is the inherent parameter condition for observing a photon blockade in the system?

(2) How does one control the driving field to achieve a strong photon blockade?

(3) How does the mechanical thermal noise affect the photon blockade?

\section{The model}

\begin{figure}[tbp]
\center
\includegraphics[bb=100 488 371 758, width=0.48 \textwidth]{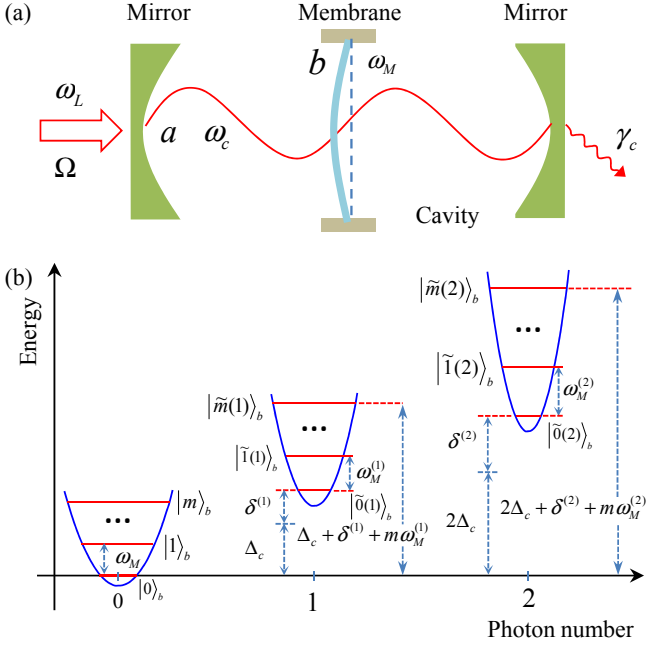}
\caption{(Color online) (a) Schematic diagram of the quadratically coupled optomechanical system with a ``membrane-in-the-middle" configuration.
(b) Diagram of the eigensystem (unscaled) of the Hamiltonian $H_{\textrm{opc}}$ (limited in the subspace spanned by the zero-, one-, and two-photon states).}
\label{setup}
\end{figure}

Specifically, we consider a quadratically coupled optomechanical system with a
``membrane-in-the-middle" configuration [see Fig.~\ref{setup}(a)]~\cite{Harris2008,Meystre2008,Harris2010,Cheung2011}.
In this setup, a thin dielectric membrane is placed at a node (or antinode) of the intracavity standing wave inside a Fabry-P\'{e}rot cavity.
The mechanical displacement of the membrane quadratically couples to the cavity photon number. In addition,
we assume that a monochromatic laser field with frequency $\omega_{L}$ is applied to weakly drive the cavity.
In a frame rotating with the driving frequency $\omega_{L}$, the Hamiltonian (with $\hbar=1$) of the system is~\cite{Harris2008}
\begin{equation}
H_{S}=\Delta_{c}a^{\dagger}a+\omega_{M}b^{\dagger }b+g_{0}a^{\dagger}a(b^{\dagger}+b)^{2}+\Omega(a^{\dagger}+a),\label{rotatingrepH}
\end{equation}
where $a$ ($a^{\dagger}$) and $b$ ($b^{\dagger}$) are, respectively,
the annihilation (creation) operators of the single-mode cavity field and the
mechanical motion of the membrane, with the respective resonant frequencies
$\omega_{c}$ ($\Delta_{c}=\omega_{c}-\omega_{L}$) and $\omega_{M}$. The third term in Eq.~(\ref{rotatingrepH})
describes the quadratic optomechanical coupling with strength $g_{0}$ between the cavity field and the mechanical
motion of the membrane.
This coupling strength $g_{0}$ should satisfy the condition $(\omega_{M}+4sg_{0})>0$ for the stability of the membrane,
where $s$ is the number of photons inside the cavity.
The last term in Eq.~(\ref{rotatingrepH}) describes the driving process, and
$\Omega$ is the driving magnitude.

Let us denote $|s\rangle_{a}$ and $|m\rangle_{b}$
($s,m=0,1,2,\ldots$) as the harmonic-oscillator number states of the cavity and the membrane, respectively;
then the eigensystem of the first three terms $H_{\textrm{opc}}=\Delta_{c}a^{\dagger}a+\omega_{M}b^{\dagger }b+g_{0}a^{\dagger}a(b^{\dagger}+b)^{2}$
in Hamiltonian~(\ref{rotatingrepH}) can be expressed as
\begin{eqnarray}
H_{\textrm{opc}}\vert s\rangle_{a}\vert\tilde{m}(s)\rangle_{b}
&=&E_{s,m}\vert s\rangle_{a}\vert\tilde{m}(s)\rangle_{b},\label{eigensys}
\end{eqnarray}
where the eigenvalues are
\begin{equation}
E_{s,m}=s\Delta_{c}+m\omega_{M}^{(s)}+\delta^{(s)}.
\end{equation}
Here we introduce the $s$-photon coupled membrane's resonant frequency $\omega_{M}^{(s)}$ and frequency shift $\delta^{(s)}$:
\begin{equation}
\omega_{M}^{(s)}=\omega_{M}\sqrt{1+\frac{4sg_{0}}{\omega_{M}}},\hspace{0.5 cm}\delta^{(s)}=\frac{1}{2}(\omega_{M}^{(s)}-\omega_{M}).
\end{equation}
The $s$-photon squeezed number state in Eq.~(\ref{eigensys}) is defined by
\begin{equation}
\vert\tilde{m}(s)\rangle_{b}=S_{b}(\eta^{(s)})\vert m\rangle_{b},
\end{equation}
where $S_{b}(\eta^{(s)})=\exp[\eta^{(s)}(b^{2}-b^{\dagger2})/2]$ is the squeezing operator, with the squeezing factor
\begin{equation}
\eta^{(s)}=\frac{1}{4}\ln\left(1+\frac{4sg_{0}}{\omega_{M}}\right).
\end{equation}
In particular, when $s=0$, we have $\vert\tilde{m}(0)\rangle_{b}=\vert m\rangle_{b}$, $\omega_{M}^{(0)}=\omega_{M}$, and $\delta^{(0)}=0$.
For convenience, the eigensystem of the Hamiltonian $H_{\textrm{opc}}$ limited in the zero-, one-, and two-photon cases is shown in
Fig.~\ref{setup}(b).
%%%%%%%%%%%%%%%%%%%%%%%%%%%%%%%%%%%%%%%%%%%%%%
%%%%%%%%%%%%%%%%%%%%%%%%%%%%%%%%%%%%%%%%%%%%%%
\section{Photon blockade in the cavity}

To investigate the photon blockade, we analytically and numerically examine the second-order correlation
function of the cavity photons.

\subsection{Approximate analytical results}

In this section, we analytically calculate the second-order correlation
function of the cavity photons by treating the weak-driving term in Hamiltonian~(\ref{rotatingrepH}) as a perturbation.
For simplicity, we phenomenologically add an anti-Hermitian term to Hamiltonian~(\ref{rotatingrepH})
to describe the dissipation of the cavity photons. The effective non-Hermitian Hamiltonian takes the form
\begin{equation}
H_{\textrm{eff}}=H_{S}-i\frac{\gamma_{c}}{2}a^{\dagger}a.\label{nonhermiteH}
\end{equation}
Here we only consider the dissipation of the cavity field and neglect the membrane's dissipation.
This approximation is justified in the time scale $1/\gamma_{c}\ll t \ll 1/\gamma_{M}$ (where $\gamma_{M}$ is the rate of
mechanical dissipation) because $\gamma_{c}\gg\gamma_{M}$. In our numerical results, the mechanical dissipation is taken into account.

In the weak-driving regime, $\Omega/\gamma_{c}\ll1$, the photon number is small, so we can work within the few-photon subspace spanned by the basis states $|0\rangle_{a}$, $|1\rangle_{a}$, and $|2\rangle_{a}$. A general state of the system in this subspace can be expressed as
\begin{eqnarray}
\vert\varphi(t)\rangle&=&\sum_{s=0}^{2}\sum_{m=0}^{\infty}C_{s,m}(t)\vert s\rangle_{a}\vert\tilde{m}(s)\rangle_{b},\label{generalstate}
\end{eqnarray}
where $C_{s,m}$ are probability amplitudes.
In terms of Eqs.~(\ref{nonhermiteH}) and (\ref{generalstate}), and the Schr\"{o}dinger equation, we obtain the equations of motion for the probability amplitudes:
\begin{subequations}
\label{equprobamplit}
\begin{align}
\dot{C}_{0,m}=&-iE_{0,m}C_{0,m}-i\Omega\sum_{n=0}^{\infty}\,_{b}\!\langle m\vert\tilde{n}(1)\rangle_{b}C_{1,n},\label{equprobamplita}\\
\dot{C}_{1,m}=&-(\gamma_{c}/2+iE_{1,m})C_{1,m}-i\Omega\sum\limits_{l=0}^{\infty }
\,_{b}\!\langle\tilde{m}(1)\vert l\rangle_{b}C_{0,l}\nonumber\\
&-i\sqrt{2}\Omega\sum\limits_{l=0}^{\infty }\,_{b}\!\langle\tilde{m}(1)\vert \tilde{l}(2)\rangle_{b}C_{2,l},\label{equprobamplitb}\\
\dot{C}_{2,m}=&-(\gamma_{c}+iE_{2,m})C_{2,m}-i\sqrt{2}\Omega\sum\limits_{n=0}^{\infty}\,_{b}\!\langle\tilde{m}(2)\vert\tilde{n}(1)\rangle_{b}C_{1,n}.
\end{align}
\end{subequations}
These transition rates can be calculated using the relations
$_{b}\!\langle\tilde{m}(s)\vert\tilde{n}(s')\rangle_{b}=\,_{b}\!\langle m\vert S_{b}(\eta^{(s')}-\eta^{(s)})\vert n\rangle_{b}$ ($s,s'=0,1,2$) and
\begin{eqnarray}
_{b}\!\langle m\vert S_{b}(\xi)\vert
n\rangle_{b}&=&\frac{\sqrt{m!n!}}{(\cosh\xi)^{n+1/2}}
\sum\limits_{l^{\prime}=0}^{\text{Floor}[\frac{m}{2}
]}\sum\limits_{l=0}^{\text{Floor}[\frac{n}{2}]}\frac{(-1)^{l^{\prime}}}{l!l^{\prime}!}\nonumber\\
&&\times\frac{(\frac{1}{2}\tanh\xi)^{l+l^{\prime}}}{(n-2l)!}(\cosh\xi)^{2l}\delta_{m-2l^{\prime},n-2l},\label{elements}\nonumber\\
\end{eqnarray}
where the function Floor$[x]$ gives the greatest integer less than or equal to $x$.

We now approximately solve Eq.~(\ref{equprobamplit}) using a
perturbation method. If there is no driving field, the cavity field will be in a vacuum.
When a weak driving field is applied to the cavity, it may excite a single photon or
two photons into the cavity, and thus we have the approximate scales
$C_{0,m}\sim 1$, $C_{1,m}\sim \Omega/\gamma_{c}$, and $C_{2,m}\sim \Omega^{2}/\gamma_{c}^{2}$.
To approximately solve Eq.~(\ref{equprobamplit}), we drop higher-order terms in the zero- and one-photon probability amplitudes,
i.e., dropping the second and third terms in Eqs.~(\ref{equprobamplita}) and~(\ref{equprobamplitb}), respectively.
For an initial empty cavity, we have $C_{1,m}(0)=0$ and $C_{2,m}(0)=0$, then
the long-time solution of Eq.~(\ref{equprobamplit}) can be approximately obtained as
\begin{subequations}
\label{gensolution}
\begin{align}
C_{0,m}(\infty)=&C_{0,m}(0)e^{-iE_{0,m}t},\\
C_{1,m}(\infty)=&-\Omega \sum\limits_{l=0}^{\infty }\frac{_{b}\!
\langle\tilde{m}(1)\vert l\rangle_{b}C_{0,l}(0)e^{-iE_{0,l}t}}{(E_{1,m}-E_{0,l}-i\frac{\gamma_{c}}{2})},\\
C_{2,m}(\infty)=&\sqrt{2}\Omega^{2}\sum\limits_{n,l=0}^{\infty
}\frac{_{b}\!\langle\tilde{m}(2)\vert \tilde{n}(1)\rangle
_{b}}{(E_{2,m}-E_{0,l}-i\gamma_{c})}\nonumber\\
&\times\frac{_{b}\!\langle\tilde{n}(1)\vert l\rangle_{b}C_{0,l}(0)e^{-iE_{0,l}t}}{(E_{1,n}-E_{0,l}-i\frac{\gamma_{c}}{2})}.
\end{align}
\end{subequations}
where $C_{0,m}(0)$ and $C_{0,l}(0)$ are determined by the initial state of the membrane. Based on Eqs.~(\ref{generalstate}) and (\ref{gensolution}),
the long-time state of the system can be obtained. Note that this approximation method has been
used to study the photon statistics in cavity QED systems~\cite{Carmichael1991,Brecha1999,Tan2002,Rice2004}.

When the cavity field is in state~(\ref{generalstate}), the equal-time (namely zero-time-delay) second-order correlation function can be written as
\begin{equation}
g^{(2)}(0)\equiv\frac{\langle a^{\dagger}a^{\dagger}aa\rangle}{\langle a^{\dagger}a\rangle^{2}}=\frac{2P_{2}}{(P_{1}+2P_{2})^{2}},
\end{equation}
where $P_{1}=\sum\limits_{m=0}^{\infty}\vert C_{1,m}(t)\vert ^{2}$ and
$P_{2}=\sum\limits_{m=0}^{\infty}\vert C_{2,m}(t)\vert ^{2}$ are the probabilities for finding a single photon and two photons in the cavity, respectively.
In the weak-driving case, we have $P_{1}\gg P_{2}$; then
$g^{(2)}(0)\approx 2P_{2}/P_{1}^{2}$.
We assume that the membrane is initially in its ground state $|0\rangle_{b}$, i.e., $C_{0,m}(0)=\delta_{m,0}$; then the long-time state of the system can be obtained from Eq.~(\ref{gensolution}). Accordingly, the photon probabilities $P_{1}$ and $P_{2}$, and the correlation function $g^{(2)}(0)$ can be obtained.

Interestingly, we examine the limit case $g_{0}/\omega _{M}\ll 1$. In
this case, we expand the squeezing operators up to zero-order in $g_{0}/\omega _{M}$. However, we keep the energy-shift
terms in the denominator of the amplitudes because these terms could be comparable to the cavity field decay rate. The correlation function can be approximated as
\begin{equation}
g^{(2)}(0)\approx\frac{4(\Delta_{c}+\delta^{(1)})^{2}+\gamma_{c}^{2}}{(2\Delta_{c}+\delta^{(2)})^{2}+\gamma_{c}^{2}}.
\end{equation}
When $g_{0}/\omega _{M}\ll 1$, the present optomechanical system reduces to an effective three-level system formed by these three states: $|0\rangle_{a}|0\rangle_{b}$, $|1\rangle_{a}|\tilde{0}(1)\rangle_{b}$, and $|2\rangle_{a}|\tilde{0}(2)\rangle_{b}$.
The corresponding values for the energy of these three states are $0$, $\Delta_{c}+\delta^{(1)}$, and $2\Delta_{c}+\delta^{(2)}$.
The difference between $\delta^{(2)}$ and $2\delta^{(1)}$ causes the energy-level anharmonicity, which is the physical origin for the appearance of the photon blockade.
Note that the driving detuning $\Delta_{c}$ is a tunable quantity by changing the driving frequency.

In the single-photon resonance (spr) case, $\Delta_{c}=-\delta^{(1)}$, the correlation function becomes
\begin{equation}
g_{\textrm{spr}}^{(2)}(0)\approx\frac{\gamma_{c}^{2}}{(\delta^{(2)}-2\delta^{(1)})^{2}+\gamma_{c}^{2}}.
\end{equation}
We have $g_{\textrm{spr}}^{(2)}(0)<1$ when $\delta^{(2)}\neq 2\delta^{(1)}$.
The larger the anharmonicity $\delta^{(2)}-2\delta^{(1)}$ is, the smaller the correlation function $g_{\textrm{spr}}^{(2)}(0)$ is. If we expand the frequency shifts $\delta^{(1)}$ and $\delta^{(2)}$ up to $g_{0}^{2}/\omega_{M}$, i.e.,
$\delta^{(1)}\approx g_{0}-g_{0}^{2}/\omega_{M}$ and $\delta^{(2)}\approx 2g_{0}-4g_{0}^{2}/\omega_{M}$, then the correlation function becomes $g_{\textrm{spr}}^{(2)}(0)\approx\gamma_{c}^{2}[4(g_{0}^{2}/\omega_{M})^{2}+\gamma_{c}^{2}]^{-1}$~\cite{Ferretti2010,Liu2013}, which is the same as that for the Kerr-type nonlinear cavity with the Kerr parameter $g_{0}^{2}/\omega_{M}$. This is because the energy spectrum of this system is the same as a Kerr nonlinearity when we expand $\delta^{(1)}$ and $\delta^{(2)}$ up to second-order in $g_{0}/\omega_{M}$.

In the two-photon resonance (tpr) case, $\Delta_{c}=-\delta^{(2)}/2$, the correlation function becomes
\begin{equation}
g_{\textrm{tpr}}^{(2)}(0)\approx\frac{(\delta^{(2)}-2\delta^{(1)})^{2}+\gamma_{c}^{2}}{\gamma_{c}^{2}}.
\end{equation}
We see $g_{\textrm{tpr}}^{(2)}(0)>1$, when $\delta^{(2)}\neq 2\delta^{(1)}$. An interesting relation is $g_{\textrm{tpr}}^{(2)}(0)g_{\textrm{spr}}^{(2)}(0)=1$.

\subsection{Numerical results}

We now turn to the numerical solution case. Including both optical and mechanical dissipations, the quantum master equation of the system is
\begin{eqnarray}
\dot{\rho}&=&i[\rho,H_{S}]+\frac{\gamma_{c}}{2}(2a\rho a^{\dagger}-a^{\dagger }a\rho -\rho a^{\dagger }a)\nonumber\\
&&+\frac{\gamma_{M}}{2}(\bar{n}_{M}+1)(2b\rho b^{\dagger}-b^{\dagger}b\rho-\rho b^{\dagger}b)\notag \\
&&+\frac{\gamma_{M}}{2}\bar{n}_{M}(2b^{\dagger}\rho b-bb^{\dagger}\rho-\rho bb^{\dagger}),\label{mastereq}
\end{eqnarray}
where we assume that the cavity field is connected with a vacuum bath, while the membrane's environment is  a heat bath at temperature $T_{M}$. $\gamma_{M}$ is the mechanical dissipation rate, and $\bar{n}_{M}=[\exp(\omega_{M}/k_{B}T_{M})-1]^{-1}$ is the average thermal phonon number of the membrane, with $k_{B}$ being the Boltzmann constant.

By numerically solving Eq.~(\ref{mastereq}), the steady state of the system can be obtained, and the second-order correlation function $g^{(2)}(0)$ of the cavity field can be calculated accordingly~\cite{numerical}. In the following, we illustrate the dependence of $g^{(2)}(0)$ on other parameters such as the coupling strength $g_{0}$, the driving detuning $\Delta_{c}$, the cavity field decay rate $\gamma_{c}$, and the thermal phonon number $\bar{n}_{M}$ in the membrane.
\begin{figure}[tbp]
\center
\includegraphics[bb=9 18 455 474, width=0.48 \textwidth]{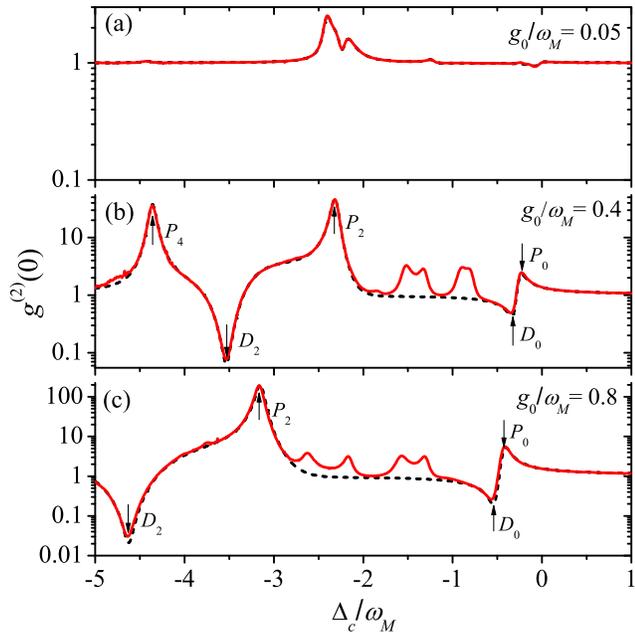}
\caption{(Color online) The equal-time second-order correlation function $g^{(2)}(0)$ versus the driving detuning $\Delta_{c}$ for various values of the optomechanical coupling strength $g_{0}$.
The solid (red) curves are numerical results using the solution of the quantum master equation~(\ref{mastereq}), while the dashed (black) curves are based on the approximate analytical solution~(\ref{gensolution}).
Other parameters are: $\gamma_{c}/\omega_{M}=0.1$, $\Omega/\omega_{M}=0.01$, $\gamma_{M}/\omega_{M}=0.001$, and $\bar{n}_{M}=0$.} \label{g2vsDelta_c}
\end{figure}

In Fig.~\ref{g2vsDelta_c}, we plot the correlation function $g^{(2)}(0)$ as a function of $\Delta_{c}$ when the coupling strength $g_{0}$ takes various values. Here, the solid (red) curves are plotted using the numerical solution of Eq.~(\ref{mastereq}), while the dashed (black) curves are based on the analytical solution in Eq.~(\ref{gensolution}).
Figure~\ref{g2vsDelta_c} shows that there is no photon blockade ($g^{(2)}(0)\ll1$) for $g_{0}<\gamma_{c}$.
When $g_{0}>\gamma_{c}$, the photon statistics transits between super-Poissonian ($g^{(2)}(0)>1$) and sub-Poissonian ($g^{(2)}(0)<1$) distributions with the change of $\Delta_{c}$. In particular, the dips and peaks in these curves correspond to the single- and two-photon resonant driving cases, respectively. In the single-photon resonant driving case, a single photon can be resonantly excited into the cavity, while the probability for finding two photons in the cavity is largely suppressed due to the energy restriction; this represents a photon blockade ($g^{(2)}(0)\ll 1$, i.e., a dip). In the two-photon resonant driving case, the probability for two photons inside the cavity is resonantly enhanced, and this corresponds to a peak in the correlation function $g^{(2)}(0)$. Therefore, the
location of these dips and peaks in Figs.~\ref{g2vsDelta_c}(b) and~\ref{g2vsDelta_c}(c) can be determined by the resonance conditions.

In the approximate analytical solution, the initial state of the membrane is assumed to be $|0\rangle_{b}$. Therefore, the well-matched
dips and peaks (marked as $D_{l=0,2}$ and $P_{l=0,2,4}$ in Fig.~\ref{g2vsDelta_c}) are determined by the resonant transitions involving the state $|0\rangle_{a}|0\rangle_{b}$, while the other peaks in the solid curves correspond to the transitions involving states $|0\rangle_{a}|n\neq0\rangle_{b}$. In particular, the dip $D_{l}$ and the peak $P_{l}$ in Figs.~\ref{g2vsDelta_c}(b) and~\ref{g2vsDelta_c}(c) correspond to the single- and two-photon resonance conditions $\Delta_{c}+\delta^{(1)}+l\omega^{(1)}_{M}=0$ and $2\Delta_{c}+\delta^{(2)}+l\omega^{(2)}_{M}=0$, respectively. The respective locations of the dip $D_{l}$ and the peak $P_{l}$ are $\Delta_{c}=-(\delta^{(1)}+l\omega^{(1)}_{M})$ and $\Delta_{c}=-(\delta^{(2)}+l\omega^{(2)}_{M})/2$. They are related to the single-photon process $|0\rangle_{a}|0\rangle_{b}\leftrightarrow|1\rangle_{a}|\tilde{l}(1)\rangle_{b}$ and the two-photon process $|0\rangle_{a}|0\rangle_{b}\leftrightarrow|2\rangle_{a}|\tilde{l}(2)\rangle_{b}$, respectively.

\begin{figure}[tbp]
\center
\includegraphics[ width=0.48 \textwidth]{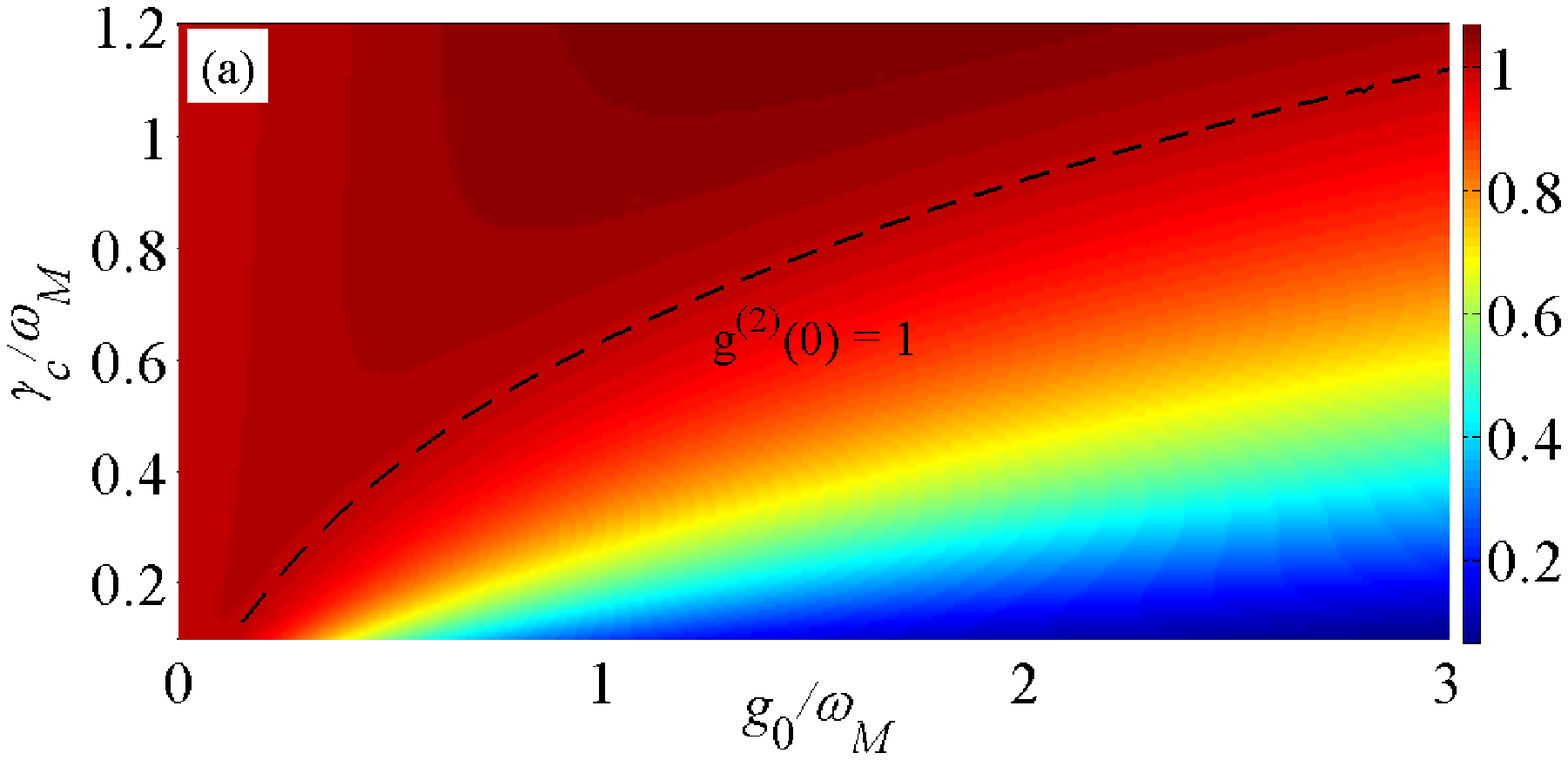}
\includegraphics[ width=0.48 \textwidth]{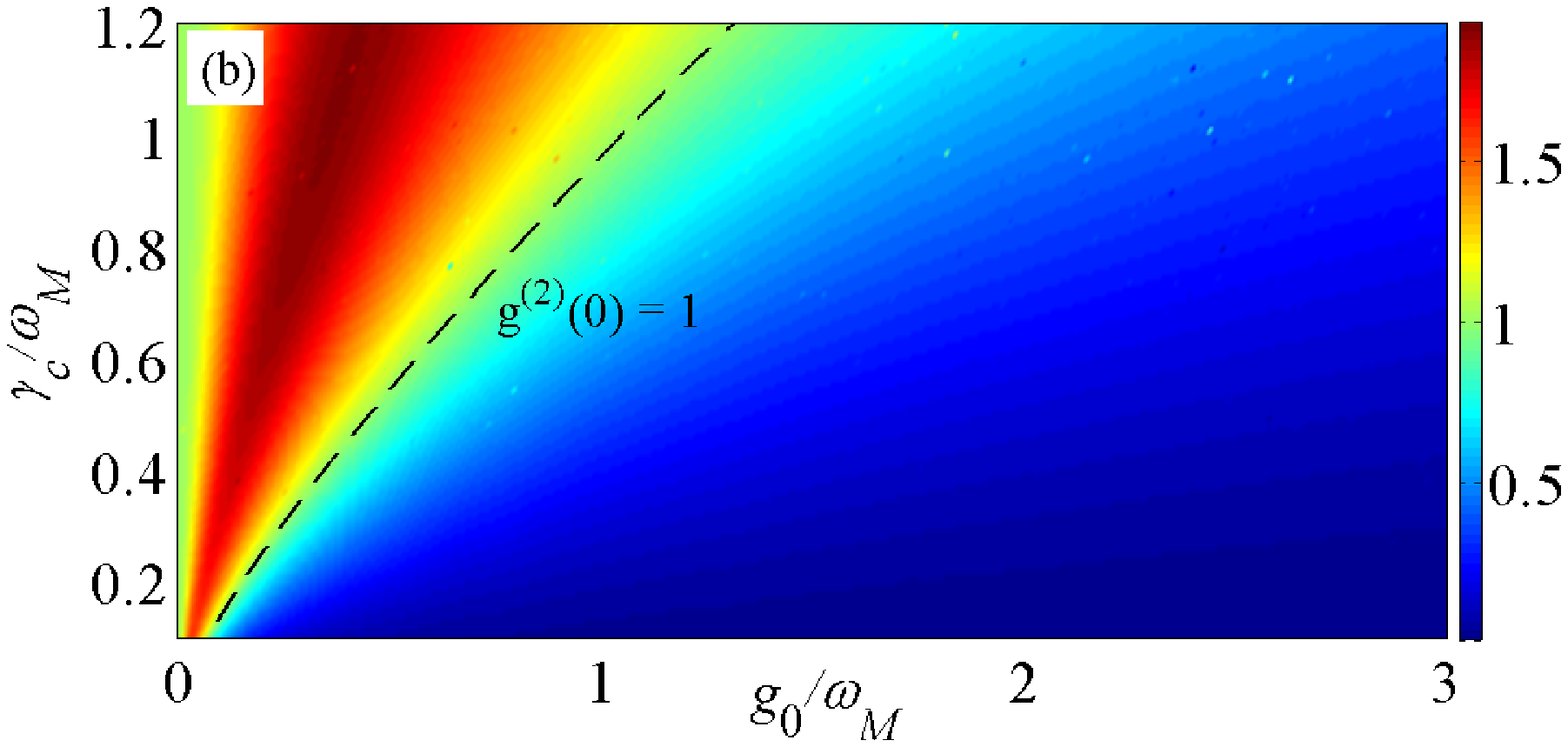}
\caption{(Color online) Plot of $g^{(2)}(0)$ as a function of the coupling strength $g_{0}$ and the cavity field decay rate $\gamma_{c}$ under the single-photon resonant driving conditions: (a) $\Delta_{c}=-\delta^{(1)}$ and (b) $\Delta_{c}=-\delta^{(1)}-2\omega^{(1)}_{M}$. Other parameters are: $\Omega/\omega_{M}=0.01$, $\gamma_{M}/\omega_{M}=0.001$, and $\bar{n}_{M}=0$.}\label{fig3}
\end{figure}

To clarify the inherent parameter condition for observing the photon blockade, we illustrate, in Fig.~\ref{fig3}, the correlation function
$g^{(2)}(0)$ as a function of $\gamma_{c}$ and $g_{0}$ under the single-photon resonant driving conditions: (a)
$\Delta_{c}=-\delta^{(1)}$ and (b) $\Delta_{c}=-(\delta^{(1)}+2\omega^{(1)}_{M})$. We can see three features from Fig.~\ref{fig3}:

(i) The curves $g^{(2)}(0)=1$ provide a boundary for different photon
distributions: super-Poissonian and sub-Poissonian. We see that the $g^{(2)}(0)\ll 1$ appears in the region $g_{0}>\gamma_{c}$.
When $g^{(2)}(0)<1$, the $g^{(2)}(0)$ decreases with increasing $g_{0}$. This means that the photon blockade grows when
increasing the quadratic coupling strength. This result is different from the linear optomechanical coupling case~\cite{Liao2013} in which the
correlation function exhibits an oscillating feature due to the modulation of the  phonon sidebands.
In the present case, when the single-photon process is resonant, the phonon sidebands will not be resonant in
the dominating two-photon transitions~\cite{explain}.

(ii) In the region $\gamma_{c}/\omega_{M}>1$, the value of $g^{(2)}(0)$ could be smaller than $1$ for a large coupling strength $g_{0}$.
This phenomenon is also different from the linear optomechanical coupling case, in which the correlation function $g^{(2)}(0)<1$ appears only
in the resolved sideband regime $\gamma_{c}/\omega_{M}<1$. We can explain this from the fact that,
though the usual resolved-sideband condition $\gamma_{c}<\omega_{M}$ does not satisfy, the two-photon coupled
phonon states $|\tilde{l}(2)\rangle_{b}$ can be still resolved in the region $\gamma_{c}>\omega_{M}$ because of
$\omega^{(2)}_{M}>\gamma_{c}$.

(iii) In the single-photon resonant driving cases, the correlation
function $g^{(2)}(0)$ for the case of $\Delta_{c}=-(\delta^{(1)}+2\omega^{(1)}_{M})$ is much smaller than that for the case of $\Delta_{c}=-\delta^{(1)}$. This phenomenon can be understood based on the photon probabilities in the cavity. As an example, we consider the parameters in Fig.~\ref{g2vsDelta_c}(c).
In the weak-driving case, the correlation function can be approximately expressed as $g^{(2)}(0)\approx 2P_{2}/P_{1}^{2}$. When the driving detuning changes from $\Delta_{c}=-\delta^{(1)}$ to $\Delta_{c}=-(\delta^{(1)}+2\omega^{(1)}_{M})$, the single-photon probability decreases by one order of magnitude, but the two-photon probability decreases by three orders of magnitude. Consequently, the driving of $\Delta_{c}=-(\delta^{(1)}+2\omega^{(1)}_{M})$ will induce a stronger photon blockade than the case of $\Delta_{c}=-\delta^{(1)}$.

Based on the above analysis, we now answer the questions (1) and (2) proposed in the introduction: (i) For observing the photon blockade ($g^{(2)}(0)\ll 1$), the system should work
in the deep-resolved-sideband regime $\gamma_{c}\ll\omega_{M}$ and the single-photon strong-coupling regime
$g_{0}>\gamma_{c}$. (ii) The single-photon phonon-sideband resonant driving $\Delta_{c}=-(\delta^{(1)}+2\omega^{(1)}_{M})$ is helpful to induce a strong photon blockade.

\begin{figure}[tbp]
\center
\includegraphics[bb=40 14 430 311, width=0.48 \textwidth]{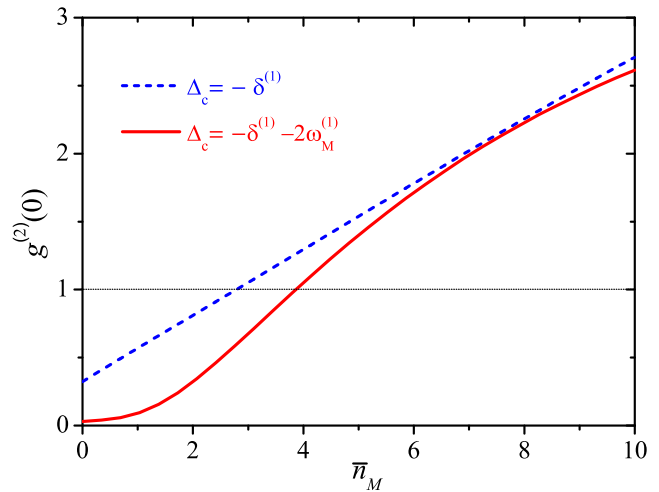}
\caption{(Color online) Plot of $g^{(2)}(0)$ versus the mechanical thermal phonon number $\bar{n}_{M}$ under the single-photon resonance conditions.
Other parameters are: $g_{0}/\omega_{M}=0.8$, $\gamma_{c}/\omega_{M}=0.1$, $\Omega/\omega_{M}=0.01$, and $\gamma_{M}/\omega_{M}=0.001$.}\label{g2vsg_and_nth}
\end{figure}

To answer question (3), we investigate the influence of the mechanical thermal phonon number $\bar{n}_{M}$ on the correlation
function, as shown in Fig.~\ref{g2vsg_and_nth}. In the single-photon resonant driving cases, the value of $g^{(2)}(0)$
increases when increasing $\bar{n}_{M}$. This means that the photon blockade is limited by the
thermal noise of the mechanical environment. To maintain the photon blockade, the mechanical thermal
noise needs to be suppressed.

\section{Conclusion and remarks}

In conclusion, we have studied the steady-state photon statistics of a quadratically-coupled optomechanical cavity, which
is weakly driven by a monochromatic laser field. We have obtained the approximate analytical expression of the second-order
correlation function for the cavity photons by treating the driving term as a perturbation. We numerically solved the quantum master equation including both optical
and mechanical dissipations. We found that the photon blockade can be induced by the quadratic optomechanical coupling
under the single-photon resonant driving condition. In particular, the phonon sideband resonant driving
$\Delta_{c}=-(\delta^{(1)}+2\omega^{(1)}_{M})$ could enhance the phonon blockade. To observe the photon blockade, the system should
work in both the deep-resolved-sideband regime and the single-photon strong-coupling regime. We have also found that the generated photon
blockade effect is limited by the thermal noise from the mechanical environment.

Finally, we present some remarks on the experimental feasibility for observing a photon blockade induced by the quadratic optomechanical couplings.
Currently, the resolved-sideband regime ($\omega_{M}\gg\gamma_{c}$) is accessible in some experimental systems.
The key challenge is the realization of $g_{0}>\gamma_{c}$. So far, the experimentally accessible couplings are too weak to reach this regime. However, recent advances have been made in the enhancement of this coupling strength. For the coupling strength $g_{0}=\frac{1}{2}\omega''_{c}(0)x^{2}_{\textrm{zpf}}$ ($x_{\textrm{zpf}}$ being the mechanical zero-point fluctuation), the value of $\omega''_{c}(0)$ has been increased significantly from about $30$ MHz/nm$^2$~\cite{Harris2010} to $20$ GHz/nm$^2$~\cite{Flowers-Jacobs2012} using a fiber cavity with a smaller mode size. For a $x_{\textrm{zpf}}\sim5$ pm, suggested in Ref.~\cite{Harris2008}, the coupling strength $g_{0}$ can reach several kilohertz. In addition, a quadratic coupling $g_{0}\sim2\pi\times0.7$ MHz has been theoretically estimated in a near-field optomechanical system~\cite{Xiao2013}.

\begin{acknowledgments}
JQL would like to thank Xin-You L\"{u} for technical support. JQL is supported by the Japan Society
for the Promotion of Science (JSPS) Foreign Postdoctoral Fellowship No. P12503. FN is partially supported by the
ARO, RIKEN iTHES Project, MURI Center for Dynamic Magneto-Optics, JSPS-RFBR Contract No. 12-02-92100, Grant-in-Aid for Scientific Research (S),
MEXT Kakenhi on Quantum Cybernetics, and the JSPS via its FIRST program.
\end{acknowledgments}


\begin{thebibliography}{99}
\bibitem{explain}      For the transitions $|0\rangle_{a}|0\rangle_{b}\leftrightarrow|1\rangle_{a}|\tilde{2n}(1)\rangle_{b}\leftrightarrow|2\rangle_{a}|\tilde{2m}(2)\rangle_{b}$, when the single-photon process is resonant, i.e., $\omega_{L}=\omega_{c}+\delta^{(1)}+2n\omega^{(1)}_{M}$, then the two-photon resonance condition is $2\delta^{(1)}+4n\omega^{(1)}_{M}=\delta^{(2)}+2m\omega^{(2)}_{M}$, which leads to $m=[(4n+1)\sqrt{1+4x}-(1+\sqrt{1+8x})/2]/(2\sqrt{1+8x})$ with $x=g_{0}/\omega_{M}$. For the dominating transitions $n,m=0,1$, there is no proper $x$ satisfying the two-photon resonant condition.
\end{thebibliography}
\end{document}